\documentclass[twocolumn,showpacs,preprintnumbers,amsmath,amssymb]{revtex4}

\usepackage{graphicx}
\usepackage{dcolumn}
\usepackage{bm}

\begin{document}

\preprint{APS/123-QED}

\title{Constraints on non-Newtonian gravity from the experiment \\ on neutron quantum states
in the Earth's
gravitational field}

\author{V.V.~Nesvizhevsky}
\affiliation{%
Institut Laue-Langevin, 6, rue Jules Horowitz BP 156, F-38042 Grenoble Cedex 9, France 
}%

\author{K.V.~Protasov}
\affiliation{
Laboratoire de Physique Subatomique et de Cosmologie, IN2P3-CNRS, UJFG,
53, Avenue des Martyrs, F-38026 Grenoble Cedex, France
}%

\date{\today}

\begin{abstract}

An upper limit to non-Newtonian 
attractive forces is obtained from the measurement of quantum states
of neutrons in the Earth's gravitational field. This limit improves the existing
contraints in the nanometer range.
\end{abstract}

\maketitle

\section{Introduction}

According to the predictions of unified gauge theories, supersymmetry, supergravity, and
string theory, there would exist a number of light and massless particles \cite{Kim}.
An exchange of
such particles between two bodies gives rise to an additional force.
Additional fundamental forces at short distances were intensively studied, in particular
during last few years following the hypothesis about ``large'' supplementary
spatial dimensions proposed by Arkami-Hamed, Dimopoulos, and Dvali
\cite{ADD,Antoniadis}, building on earlier ideas in Refs \cite{RS,Visser,Anton,Lykken}.
For a review of theoretical works and recent
experimental results, see \cite{PDG,Bordag,Onofrio,LCP}.
This hypothesis could be verified using neutrons because the absence of
an electric charge allows one to strongly suppress the false electromagnetic effects
\cite{Isaker}. It was noticed in \cite{BN} that the measurement of the neutron
quantum states in the Earth's gravitational field \cite{PRD2003} is sensitive
to such extra forces in the sub-micrometer range. In the case of $n=3$ extra
dimensions, the characteristic range is just in the nanometer domain
\cite{ADD,Isaker} which is accesible in this experiment.
The first attempt to establish a model-dependent boundary in the range 1--10 $\mu$m,
was presented in ref. \cite{Abele}.

An effective gravitational interaction in presence of an additional
Yukawa-type force is parametrized as:
\begin{eqnarray}
\label{intergen}
V_{\mbox{\tiny eff}}(r)= G \frac{m_1 m_2}{r}\left(1 + \alpha_G \mbox{e}^{-r/\lambda}
\right).
\end{eqnarray}
Here, $G$ is the Newtonian gravitational constant, $m_1$ and $m_2$ are
interacting masses, $r$ their relative distance, $\alpha_G$
and $\lambda$ are strength and characteristic range of this hypothetical
interaction.

The experiment \cite{PRD2003} consists in the measurement of the neutron flux through
a slit between a horizontal mirror on bottom and a scatterer/absorber on top as a
function of the slit size $\Delta h$.
 This dependence is sensitive to
the presence of quantum states of neutrons in 
the potential well formed by the Earth's gravitational field and the mirror. In particular,
the neutron flux was measured to be equal to zero within the experimental accuracy
if the slit size $\Delta h$ was smaller than the
characteristic spatial size (a quasiclasssical turning point height)
of the lowest quantum state of $\sim 15 \mu$m in this potential well.
The neutron flux at the slit size $\Delta h < 10 \mu$m was lower by at least a factor
of 200 than that for the lowest quantum state($\Delta h \approx 20 \mu$m).
If an additional short-range force of sufficiently high strength
would act between neutrons and the mirror then it would modify
the quantum states parameters: an attractive force would ``compress'' the
wave functions towards the mirror, while a repulsive force would shift them up. 
In this experiment, no deviation from the expected values
was observed within the experimental accuracy.
This accuracy is defined by the uncertainty in the slit size which
can be conservatively estimated as $\approx $ 30 \% for the lowest quantum state
\cite{PRD2003}.


The motion of neutrons in this system over the vertical axis $z$ could
be considered, within first and quite good approximation, as an
one dimentional problem for which the mirror provides an
infinitely high potential. The interaction between neutrons and the Earth
is described by the first term in equation (\ref{intergen}) and can be
approximated by the usual linear potential ($r=R+z$):
\begin{eqnarray}
\label{mgz}
V(z)= mgz
\end{eqnarray}
with $g = GM/R^2$, $R$ being the Earth's radius, $M$ its mass, $m$
the neutron mass.

The second term in equation (\ref{intergen}) introduces an additional
interaction. Due to the short range of this interaction, its main contribution
is provided by the interaction of neutrons
with a thin surface layer of the mirror and the scatterer.

In this article, we estimate an upper limit on an additional
attractive short-range force, which could be established from this
experiment in a model-independent way.
We show that it could not be significantly improved in
any more sophisticated model.

\section{Attractive interaction}
Let us estimate firstly the interaction of neutrons with the mirror.
If the mirror's density is constant and equals to $\rho_m$, then an additional
potential of the interaction between neutrons and the mirror is given by:
\begin{eqnarray}
V'(z') &=& - G \alpha_G m \rho_m \nonumber \\
&&\int_{\mbox{\tiny mirror}} d^3r
\frac{\exp \left(-\sqrt{x^2+y^2+(z-z')^2}/\lambda\right)}{\sqrt{x^2+y^2+(z-z')^2}}.
\end{eqnarray}
The volume integral is calculated over the mirror bulk:
$- \infty < x,y < \infty$, $z <0$ (in fact, over the neutron's vicinity with the
size of the order of a few $\lambda$
due to the exponential convergency of these integrals). It can be calculated
analytically for small $\lambda$:
\begin{eqnarray}
\label{newint}
V'(z)=- U_0 \mbox{e}^{-z/\lambda}
\end{eqnarray}
with $U_0 = 2\pi G \alpha_G m \rho_m \lambda^2$.

\begin{figure}[h!]
\begin{center}
\includegraphics{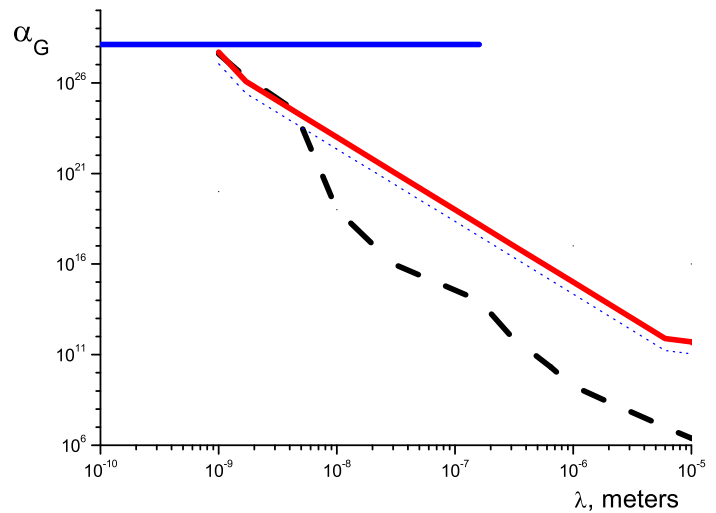}
\end{center}
\caption{
The constraints on $\alpha_G$ following from the
experiment \cite{PRD2003} (the solid line) in comparison with that from the measurement of
the Casimir and the van der Waals forces \cite{Bordag} (the dashed line).
The dotted line shows a limit which can be easily obtained by an improvement
of this experiment.
The solid horizontal line represents the limit 
established from the atomic experiment \cite{Asakusa}.
\label{figlimit}}
\end{figure}

The simplest upper limit on the strength of an
additional interaction follows from the condition 
that this additional interaction does not create itself
any bound state. It is known \cite{MQExo} that for an exponential
attractive ($\alpha_G > 0$) potential (\ref{newint}) this means that:
\begin{eqnarray}
\label{limsimple}
\frac{U_0m \lambda^2}{\hbar^2}  < 0.72 .
\end{eqnarray}
This condition gives a boundary for an additional
potential strength:
\begin{eqnarray}
\label{statesbound}
\alpha_G = 0.72 \frac{2}{\pi} \frac{\rho}{\rho_m} \frac{\hbar}{mg\lambda^2}
\frac{\hbar}{m\lambda} \frac{R}{\lambda},
\end{eqnarray}
$\rho$ being the Earth's averaged density. In this experiment, both densities
are close to each other $\rho \approx \rho_m$, therefore their ratio
$\rho/\rho_m$ is close to 1. However an adequate choice of the mirror material (coating) would
easily allow one to gain a factor of 3--5 in the sensitivity in future experiments.
One obtains the following numerical boundary:
\begin{eqnarray}
\label{statesnumbound}
\alpha_G = 1. \cdot 10^{15} \left( \frac{1 \mu\mbox{m}}{\lambda}\right)^4
\end{eqnarray}
Here, 1 $\mu$m is chosen as a natural scale for this experiment.
This limit
is presented in Figure~\ref{figlimit} in comparison with the limits coming
from the Casimir-like and van der Waals force measurement experiments \cite{Bordag}.
One can note that, in the realistic case, one has to establish
a condition of non existence of an additional bound state for the sum of (\ref{mgz}) and
(\ref{newint}) but not for the interaction (\ref{newint}) alone.
A presence of the linear potential modifies slightly the critical value
in (\ref{limsimple}). For instance, for $\lambda = 1 \mu$m
it equals approximately to 1.0 and for $\lambda = 0.1 \mu$m
it equals to 0.74. For smaller $\lambda$, this value is evident to
tend to 0.72. It is possible to explain qualitatively why the
strength of an additional interaction should be higher in presence of the $mgz$-potential
than without it. When a bound state just appears, then its wave function is
extremely spread. If one adds a supplementary ``external'' confining
potential, it does not allow the wave function to be spreaded and thus
one needs a stronger potential to create a bound state.
The range of presented $\lambda$ is 1 nm -- 10 $\mu$m.
A deviation from a straight line in the solid curve at 1 nm is due to the finite range
of increase of the mirror effective nuclear potential (impurities on the surface and
its roughness). The same effect at $\lambda \approx 10 \mu$m is due to an ``interference'' of
the potentials (\ref{mgz}) and (\ref{newint}).

It is interesting to compare this analytical limit (\ref{statesnumbound})
to an analogous
expression obtained in \cite{Bordag,Onofrio} for the Casimir force-like experiments.
The simplest boundary $\alpha = \alpha(\lambda)$ following from these experiments is given by
a formula:
\begin{eqnarray}
\label{cas}
\alpha_C = C_C \frac{\exp(d_0/\lambda)}{\lambda^3}.
\end{eqnarray}
Here $d_0$ is a gap separation and $C_C$ is a constant depending on the
geometry of different experiments. This function increases exponentially when
$\lambda$ tends to zero. This behaviour is clearly seen in Figure~\ref{figlimit}.
In the experiment \cite{PRD2003} we obtain
\begin{eqnarray}
\label{neut}
\alpha_n = \frac{C_n}{\lambda^4}
\end{eqnarray}
and $\alpha_n$ increases only as $1/\lambda^4$!
This difference between (\ref{cas}) and (\ref{neut}) means that,
in principle, for any $C_C$ and
$C_n$, one could find a domain of sufficiently small 
but experimentally observable $\lambda$ in which the limit
obtained from our experiment is stronger than that obtained from a Casimir
force-like experiment.

Up to now, we have not discussed the scatterer. As we can see from Figure~\ref{figlimit},
a competitive limit could be established only for $\lambda$ much smaller than the scatterer
roughness amplitude of the order of 1 $\mu$m. Therefore an influence of the scatterer
is negligible.

\section{Repulsive interaction}

Unfortunately, this experiment does not allow us to establish a competitive limit for a
repulsive interaction. In this case, there could be no ``additional'' bound state. Here, 
instead of the condition of ``non-existence'' of a bound state, one could consider
the critical slit size for which the first bound state appears in this
system. Such an approach would be model-dependent due
to uncertainties in the description of the interaction of neutrons with the scatterer.
Nevertheless, it is possible to obtain a simple analytical expression 
for small $\lambda$ and to show explicitly a difference
in sensitivity of this experiment to an attractive and to a repulsive additional
interactions.

Let us remind the reader that the quasiclassical approximation for the potential composed
by the linear term $mgz$ and by the infinite mirror potential describes the problem very well.
For instance, it gives the ground state energy,
for which the quasiclassical approximation is usually expected to be not valid,
within 1 \% \cite{MQExo,Flugge}. The Bohr-Sommerfeld quantification condition in this
case assumes that the quasiclassical wave function
\begin{eqnarray}
\Psi_{\mbox{\tiny qc}}(z) = \frac{C_1}{\sqrt{p(z)}} \sin
\left(  \frac{1}{\hbar} \int_z^b p(z') dz' + \frac{\pi}{4} \right),
\end{eqnarray}
with $p(z)=\sqrt{2m(E-mgz)}$  ($0 < z < b$, $b$ being the turning point) equals to zero at $z=0$ (at
the inpenetrable mirror). This condition, for $n$-th quantum state, means
\begin{eqnarray}
\frac{1}{\hbar} \int_0^b p(z) dz + \frac{\pi}{4} =\pi (n+1), \hspace{1cm} n=0,1, \ldots
\end{eqnarray}
and the corresponding energies are:
\begin{eqnarray}
E_n = \left(\frac{9\pi^2}{8}\right)^{1/3} (mg^2\hbar^2)^{1/3} (n+3/4)^{2/3}.
\end{eqnarray}
With an additional short-range repulsive interaction $V'(z)=U_0 \exp \{-z/\lambda\}$,
the problem could be solved in an analogous way. One can choose
such $z_0$ that $\lambda \ll z_0 \ll b$. For $0 < z < z_0$, the $mgz$
potential is negligible and the problem can be solved analytically \cite{MQExo}. For
$z_0 < z <b$, one can use the quasiclassical
approximation. At $z = z_0$ the quasiclassical wave function should be equal to
the exact one:
\begin{eqnarray}
\Psi_{\mbox{\tiny ex}}(z_0) = C_2 \sin (kz_0 + \delta_0),
\end{eqnarray}
with the phase shift
\begin{eqnarray}
\delta_0= -4k\lambda \left[ C + \ln \frac{\xi}{2}+ \frac{K_0(\xi)}{I_0(\xi)}\right].
\end{eqnarray}
Here  $k=\sqrt{2mE/\hbar^2}$, $\xi=\sqrt{8mU_0\lambda^2/\hbar^2}$ , $C=0.577$ is the
Euler constant, $I_0(\xi)$ and $K_0(\xi)$ are the modified Bessel functions of
pure imaginary argument. This phase shift modifies the
Bohr-Sommerfeld quantification condition and introduces a shift $\delta E_n$ in the
level's energy. If this shift is small with respect to the nonperturbated value, one obtains the
following relation between the energy shift $\delta E_n$ and the phase shift $\delta_0$: 
\begin{eqnarray}
\frac{1}{\hbar} 
\left[ \frac{2E_n^3}{mg^2}\right]^{1/2} =\delta_0 .
\end{eqnarray}
For very strong repulsive potential, $\xi \gg1$ , the phase shift is equal to
\begin{eqnarray}
\delta_0 \approx -4k\lambda \ln \frac{\xi}{2}.
\end{eqnarray}
An experimental upper limit on the energy shift $\delta E_n$ would impose
an upper limit on  $\alpha_G$ for a repulsive interaction:
\begin{eqnarray}
\label{repuls}
\frac{2U_0m\lambda^2}{\hbar^2}  < \exp (\lambda_0/\lambda).
\end{eqnarray}
with
\begin{eqnarray}
\lambda_0=  \frac{\delta E_n}{mg}.
\end{eqnarray}
 or
\begin{eqnarray}
\label{repulsalph}
\alpha_G \leq \frac{1}{\pi} \frac{\hbar}{mg\lambda^2}
\frac{\hbar}{m\lambda} \frac{R}{\lambda} \exp (\lambda_0/\lambda).
\end{eqnarray}
Direct comparison of relation (\ref{repulsalph})
to (\ref{limsimple}) and (\ref{statesbound}) shows that 
the limit (\ref{repulsalph}) at small $\lambda$ is sufficiently less restrictive
than that for an attractive one (\ref{statesbound}) due
to the exponential factor.

Note that the presented approach could be applied as well for an 
attractive interaction $V'(z)=-U_0 \exp \{-z/\lambda\}$. In this case, the phase
shift is equal to
\begin{eqnarray}
\delta_0= -4k\lambda \left[ C + \ln \frac{\xi}{2}- \frac{\pi}{2}
\frac{N_0(\xi)}{J_0(\xi)}\right].
\end{eqnarray}
Here $J_0(\xi)$ and $N_0(\xi)$ are the Bessel functions.
This expression is maximal in the vicinity of the first Bessel function's
zero $\xi_0 \approx 2.40$ where it can be developped as
\begin{eqnarray}
\delta_0 \approx -4k\lambda \frac{\pi}{2} \frac{N_0(\xi_0)}{J_1(\xi_0)}\frac{1}{\xi-\xi_0}.
\end{eqnarray}
One thus obtains the relation between the strength of an attractive potential and
the energy shift $\delta E_n<0$:
\begin{eqnarray}
\xi \leq \xi_0- \pi\frac{N_0(\xi_0)}{J_1(\xi_0)}\frac{\lambda}{\lambda_0}.
\end{eqnarray}
with  $\lambda_0=  \frac{|\delta E_n|}{mg}$, $N_0(\xi_0)\approx 0,51$,  $J_1(\xi_0)\approx 0,52$.
If one omits the second term in the latter equation, one obtains exactly the
relations (\ref{limsimple}) and (\ref{statesbound}) representing a condition of
non-existance of an additional bound state.

Let us mention that this formula allows us to estimate the capability of this
experiment to further improve an upper limit on $\alpha_G$ in the case of an
attractive interaction:
in order to improve significantly (by an order of magnitude) the limit
(\ref{statesbound}) in a given range of $\lambda$ it is necessary to determine  
experimentally and theoretically the parameters of quantum states with a precision
$\lambda_0/\lambda \approx 1$.

\section{Occupation numbers}
The considerations presented above are valid only if the neutron population in the lowest
quantum state in such a system (with an additional interaction included) if sufficiently high
to provide a measurable signal/noise ratio. As we have mentioned in the Introduction,
the experiment \cite{PRD2003} would allow one to identify an additional quantum state
if its occupation number
would not be suppressed by more than a factor of 200 compared to that for other states.

In order to calculate the occupation numbers, let us start with a general expression for the
probability of a rapid transition from a state $k$ with the wave function $\Psi_k(x)$
to a state $n$ with the wave function $\Phi_n(x)$ which is given by a formula \cite{LL}:
\begin{eqnarray}
w_{k \rightarrow n}= \left| \int \Psi_k(x) \Phi_n^*(x) dx \right|^2.
\end{eqnarray}
For a few initial quantum states, the probability $w_n$ is a
sum (an integral for continius spectrum) over them:
\begin{eqnarray}
\label{fewtran}
w_n = \sum_k f_k w_{k \rightarrow n}.
\end{eqnarray}
with the occupation numbers $f_k$ of initial states.

To obtain an analytical expression for the occupation numbers, let us consider
a simplified model of a harmonic oscillator in a final state and a plane wave
in an initial one. An explicite analytical shape of the final state wave function
does not play a role (the only important parameter is its spatial size)
and would not modify considerably the occupation numbers.
The used wave functions are equal to  
\begin{eqnarray}
\Phi_n(x) = \frac{1}{\sqrt[4]{\pi x_0^2}} \frac{1}{\sqrt{2^n n!}}
\exp \left(-\frac{x^2}{2x_0^2}\right)
H_n\left( \frac{x}{x_0}\right).
\end{eqnarray}
and
\begin{eqnarray}
\Psi_k(x) = \frac{1}{\sqrt{2\pi}} \mbox{e}^{ikx}
\end{eqnarray}
Here $H_n(z)$ are the Hermite polynomials and $x_0$ represents a characteristic geometric
size of the problem.

If initial states are populated according to the Gaussian law with a chercteristic
momentum $k_0$ then
\begin{eqnarray}
f_k = \exp \left(-\frac{k^2}{k_0^2}\right)
\end{eqnarray}
and all integrals  (\ref{fewtran}) can be calculated analytically \cite{PBM}.
The occupation numbers of the final states 
appear to be equal to:
\begin{eqnarray}
w_n &=& \frac{k_0x_0}{\sqrt{1+ (k_0x_0)^2}}
\left(\frac{(k_0x_0)^2-1}{(k_0x_0)^2+1}\right)^{n/2} \nonumber\\
&& \times P_n\left(\frac{(k_0x_0)^2}{\sqrt{(k_0x_0)^4-1}}\right).
\end{eqnarray}
Here $P_n(z)$ is the Legendre polinomial.
For instance, for the lowest states with $n=0$ and $n=1$:
\begin{eqnarray}
w_0 = \frac{k_0x_0}{\sqrt{1+ (k_0x_0)^2}}; \hspace{2mm} w_1=w_0^3.
\end{eqnarray}
If $k_0x_0\gg 1$ then the occupation numbers are approximately equal
for all states:
\begin{eqnarray}
w_n \approx 1.
\end{eqnarray}

Let us apply these expressions to our problem.
For the gravitational quantum states,  $x_0 \approx 6 \mu$m;
the vertical velocity distribution has a characteristic velocity of $v_0\approx 50$~cm/s. For
these states, $k_0x_0 \approx 50 \gg 1$ and all states should have approximately
the same occupation numbers.

If an additional bound state were created by the interaction (\ref{newint})
then the characteristic size of such a state should be of the order 
of the range of the interaction $\lambda$ (or bigger). 
For the interaction range, for which this experiment establishes a competitive
limit, one obtains $w \approx k_0 \lambda \approx 0.1$
for $\lambda = 10$~nm and $w \approx k_0 \lambda \approx 0.01$ for $\lambda = 1$~nm.
As we mentioned previously, if such a state exists it would be detected in this experiment.

Moreover, these numbers represent a lower estimation for the occupation numbers
because they assume the state to be deeply bound. For a just appearing state, the
characteristic size $x_0$ would be sufficiently bigger than the range of the interaction
and would be close to that of the unperturbed states in the gravitational potential.
Thus the population would not be almost suppressed.

\section{A limit from exotic atoms}
To complete our analysis in the nanometer range, let us note that
a competitive limit follows from a recent experiment with 
antiprotonic atoms as well. The idea of such an analysis is the same as 
in ref. \cite{BMS}.
An additional force between a nucleus of mass $M$ and an antiproton would
change the spectrum of such an atom. The effective orbit radius $r_0$ for usually
studied antiproton-nucleus atoms is about a few hundreeds fm.
Therefore, for the range of a hypothetical Yukawa force $\lambda \gg r_0$, an additional
potential would have a $1/r$ behaviour and would change the Coulomb potential:
\begin{eqnarray}
Ze^2 \rightarrow Ze^2 + mMG \alpha_G
\end{eqnarray}
The energy spectrum $E_n = - \frac{(Ze^2)^2}{2n^2} \mu$
of an atom with a reduced mass $\mu$ as well as the transition
frequences $\nu_n$ would be modified as well:
\begin{eqnarray}
\nu_n \rightarrow \nu_n \left(1 + \frac{2mMG \alpha_G}{Ze^2}\right) \equiv
\nu_n  + \delta \nu_n.
\end{eqnarray}
Thus, the constant $\alpha_G$ is related to the shift $\delta \nu_n$ in the transition
frequencies by a formula
\begin{eqnarray}
\label{alpvisfreq}
\alpha_G = \frac{Ze^2}{2mMG}\frac{\delta \nu_n}{\nu_n} 
\end{eqnarray}
An upper experimental limit on the transition frequency shift $|\delta \nu_n|$
imposes an upper limit on $|\alpha_G|$. The most precise measurement
of the energy spectrum of antiprotonic atoms was
done for \={p}$^3$He$^+$ and \={p}$^4$He$^+$ atoms by the ASAKUSA collaboration
at the Antiproton
Decelerator at CERN \cite{Asakusa}. In this experiment, 13 electromagnetic transitions
between different levels of these atoms were investigated. No deviation was found from
the values expected within the QED calculations \cite{Korobov}.
Equation (\ref{alpvisfreq}) allows $\alpha_G$ to be extracted.
The averaged over these 13 transitions value of $\alpha_G$ is equal:
\begin{eqnarray}
\alpha_G = (3 \pm 10) \cdot 10^{27},
\end{eqnarray}
and is compatible with zero. 
A $1\sigma$ upper limit on $|\alpha_G|$ from this experiment is:
\begin{eqnarray}
|\alpha_G| \leq 1.3 \cdot 10^{28}.
\end{eqnarray}
This value is presented by the solid horizontal line in Figure~\ref{figlimit}.

\section{Conclusions and perspectives}
An upper limit to an additional attractive force is established from the measurement
of quantum states
of neutrons in the Earth's gravitational field. 
Relatively high sensitivity of the experiment \cite{PRD2003} to a hypothetical
additional force is due to the following factors: 
firstly, no ``background'' electromagnetic interactions;
secondly, the characteristic size of the neutron
wave function in the quantum states fits well to the range of interest for
the short-range forces; 
finally, non-negligible probability to find neutrons
(quantum-mechanical object) at distances much closer to the mirror than 
the average value of 10 $\mu$m. 

The limit (\ref{statesnumbound}) improves the existing
contraints \cite{Bordag} in the nanometer range even if this
experiment was neither conceived nor optimized to establish this limit.

However, it can be easily improved in the same kind of experiment with
some evident modifications, for instance, one can choose a mirror material (coating) with
higher density.
A more significant gain in the sensitivity could be
achieved in a dedicated neutron experiment which will be presented in a forthcoming paper.

\bigskip

\begin{acknowledgments}
We are very grateful to P.~van Isaker, R.~Onofrio, V.A.~Rubakov,
M.E.~Shaposhnikov, and P.G.~Tilyakov for advice and useful discussions.
\end{acknowledgments}


\end{document}